# SNIF: a Self-Navigator Intelligent Filter correcting off-center and spin-history artifacts in quasi-random radial MRI

**Tanguy Boucneau[1], Brice Fernandez[2], Peder Larson[3], Luc Darrasse[1], and Xavier Maître[1]✉**

**For self-navigation and self-gating in MRI, self-navigators are extracted from the complex values of the k-space center, called DC-signal, through a temporal frequency analysis. However, this signal often carries artifacts. This paper introduces a Self-Navigator Intelligent Filter (SNIF) that models the physical origin of the k-space center artifacts to properly estimate them and thus more accurately remove them. The artifact affecting the DC-signal $S_{DC}$ generally has two main origins: the jitter of the k-space center ("off-center artifact") and the cumulative history of the spins ("spin-history artifact"). Both add noise-like artifacts to the signal. The off-center effect can be modeled as a perturbative signal $S_0$ and the spin-history effect as a convolution kernel $h$ so that the measured DC-signal is $S_{DC} = S_m + S_0 \star h$, where $S_m$ is the motion-related component which effectively contains the expected self-navigators. Methods to compute $h$ and $S_0$ were developed, which gives the possibility to extract $S_m$ from $S_{DC}$. They were implemented and evaluated on thoracic UTE datasets acquired on a healthy volunteer. Qualitative visual comparison and quantitative spectral analysis of the DC-signal before and after SNIF demonstrate that the method improves both the accuracy and the precision of the extracted respiratory and cardiac self-navigators. By considering the physical origin of the DC-signal artifacts, SNIF makes both respiratory and cardiac self-navigators more sensitive to motion, which supports finer MRI motion tracking.**

Standard $^1$H MRI of the lung is particularly challenging for three main reasons [1–3]: the low pulmonary mass density, the short $T_2^*$ of lung tissues and the physiologic motions. Ultrashort echo-time (UTE) pulse sequences have been shown to address the last two main problems [4–7]. In particular, as the k-space sampling is radially performed, the DC-signal is sampled every repetition and time-resolved information about the motion state of the lung and/or the heart can usually be extracted out of the corresponding 1D signal. This signal is usually referred to as a self-navigator [8,9].

In most studies, two steps are implemented to extract the self-navigators from the DC-signal: i) separation of respiratory and cardiac information and ii) filtering of high-frequency noise. The latter noise arises when a quasi-random choice of spoke orientations is selected for k-space sampling (computationally-random [10] or golden-angle based [11,12]) and gradient spoiling is defective. Both steps are usually performed through a temporal frequency analysis because respiratory and cardiac motions can be approximated by quasi-periodic motions with different frequencies. In practice, this consists of band-pass filtering the DC-signal to separate the information related to each motion and to reduce the background noise [10–15]. It holds as long as the physiologic motions are quasi-periodic, which is not always the case over a several-minute acquisition, even in healthy subjects. Moreover, as the filtering is based only on the assumed frequency content of the signal, the origin and the nature of the high-frequency noise are not integrated in the correction process.

In this paper, we introduce SNIF, a Self-Navigator Intelligent Filter designed to suppress the apparent high-frequency noise by considering its physical origin in the context of 3D-UTE with radial sampling in random order. SNIF assumes that this apparent noise results from random noise accumulated along the measurement chain and two different physical artifacts:

1. an off-center artifact as the DC-signal measurement is not performed exactly at the center of k-space because of overall system

[1] Université Paris-Saclay, CEA, CNRS, Inserm, BioMaps, Orsay, France

[2] Applications & Workflow, GE Healthcare, Orsay, France

[3] Department of Radiology and Biomedical Imaging, University of California San Francisco, San Francisco, CA, USA

✉ Email: xavier.maitre@universite-paris-saclay.fr

imperfections ($B_0$ inhomogeneity, gradient delays…), and
2. a spin-history artifact as spurious contributions from previous repetitions may contaminate the current transverse magnetization state.

By removing both contributions out of the signal, we aim at extracting more reliable self-navigators. Compared with common techniques, we expect those corrected self-navigators to better reflect the underlying motion fields in the patient's thorax and be less dependent on acquisition biases. In turn, this may improve the quality of retrospective motion correction.

## Theory

### Modelling of the off-center and spin-history artifacts

Let $S_{DC}$ be the DC-signal measured at the expected center of k-space, recorded by a given reception coil element in the course of a radial MR pulse sequence. It corresponds to the first sampled data point along each center-out radial readout. Assuming a motionless state, we write this signal $S_0(t)$. If this measurement were performed at the theoretical center of k-space with no added noise source, $S_0(t)$ would remain constant and equal to the mean intensity of the image acquired with each coil element $S_0$.

Because of hardware and pulse-sequence imperfections, the first data point is, in practice, never sampled exactly at the theoretical center of k-space. This leads to an off-center artifact characterized by a temporal variation of the recorded $S_0$ value, which depends on both the readout gradient direction and the subject's morphology. Moreover, in the general case, the measured signal also depends on cumulative former transverse magnetization states reached in preceding sequence repetitions. It introduces a spin-history effect in the recorded $S_{DC}$. This effect was modelled by a convolution kernel $h = [h_{N_{Rep}} \dots h_3\, h_2\, h_1\, h_0]$, where $N_{Rep}$ is the number of relevant preceding sequence repetitions which influence the current measurement. Each one lasts $TR$ and $h_0$ corresponds to the current repetition.

Thereby, the DC-signal is a sum of the back-shifted and re-weighted $S_0$ signal:

$$S_{DC} = S_0 \star \mathrm{h} \qquad [1]$$

where $\star$ is the discrete convolution operator. In other words, the effect of previous magnetization states on the current DC-signal is here assumed to be linearly dependent on $S_0$ values recorded before. Considering low frequency motion from respiration and cardiac contraction, the DC-signal can be modelled as a small additive contribution to $S_0$, written $S_m$. The DC-signal $S_{DC}$ becomes therefore:

$$S_{DC} = (S_m + S_0) \star h \qquad [2]$$

For a quasi-random off-centered radial sampling of k-space, which is the case when the entire k-space is uniformly sampled with an order of spokes ruled by randomness or a law that aims at maximizing the uniformity of k-space sampling within any small number of successive spokes, $S_0$ can be considered as a noise containing information within a broad range of frequencies. In comparison, if the motion has a time evolution characterized by a quasi-period $T \gg N_{\text{rep}}TR$ (in the case of a quasi-periodic motion), the motion state does not change much from one repetition to the other. This inequality is experimentally verified for thoracic motions for most short $TR$ pulse sequences ($TR < 10$ ms) because the spin-history effect depends mainly on a small number of previous repetitions ($h$ can be characterized by a small number of values, namely $N_{Rep}$ is small, as demonstrated later). For that reason, [2] can be simplified as

$$S_{DC} = S_m + S_0 \star h \qquad [3]$$

where $S_0 \star \mathrm{h}$ constitutes an additive pseudo-noise for the rescaled self-navigators $S_m$.

### Estimation of the off-center and spin-history artifacts

Every spoke of a 3D 'kooshball' trajectory is characterized by two angles in spherical coordinates: a polar angle, $\theta$, and an azimuthal angle, $\varphi$ (Figure 1a). $S_{DC}$ is processed in Matlab® to extract $\hat{S}_0$, an estimation of the k-space off-center artifact $S_0$, with a method based on the $C^1$-constrained interpolation of k-space data from the time domain to a $(\theta, \varphi)$ representation of k-space (Figure 1b). Then, data are filtered by applying successively median, sliding window, and Gaussian low-pass filters before being interpolated back from $(\theta, \varphi)$ into the time domain (Figure 1c-d).

The filters were applied in this specific order. Their characteristic parameters were tuned manually to remove the noise without degrading the pattern varying slowly along $\theta$ and $\varphi$. This manual tuning process was performed on the data acquired by several representative reception channels. For the specific sets of acquisition parameters here, it led to a median filter kernel of size 3×3, a sliding window width ranging from 1 (for $\theta$ values around $\pi/2$) to 61 (for $\theta$ values close to 0 and $\pi$), and a Gaussian filter kernel of size 21 and standard deviation 5. The size of the discretized $(\theta, \varphi)$ space in which these filters were applied was 96×192. The same parameters were then used for the entire dataset. Note that this method can be easily transposed to 2D imaging by interpolating data in the polar representation of k-space along $\varphi$ only.

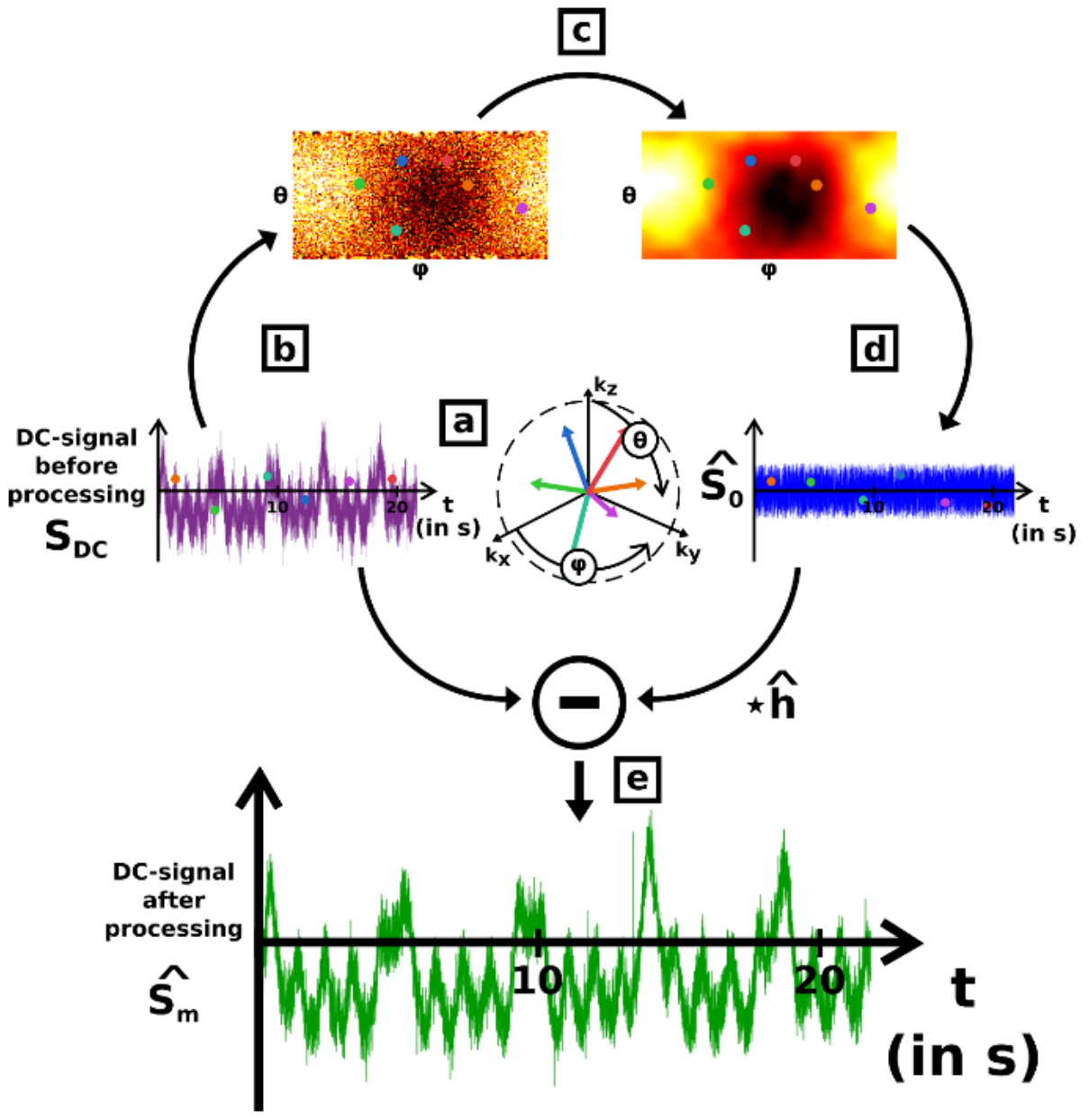

**Figure 1:** SNIF extraction of the off-center component $\hat{S}_0$ from $S_{DC}$ and its role in the correction process: a) Assignment of a pair of angular coordinates $(\theta, \varphi)$ to each DC-signal data point; b) Interpolation and regridding of the DC-signal along $\theta$ and $\varphi$: motion and spin-history components add apparent noise in this matrix and the off-center component adds a slowly varying pattern along $\theta$ and $\varphi$; c) Smoothing of the resulting matrix to keep only the off-center component; d) Inverse interpolation process to extract the $\hat{S}_0$ signal: motion information and spin-history artifact are removed; e) Filtering of $S_{DC}$ with SNIF: $\hat{h}$ is estimated from $S_{DC}$ and $\hat{S}_0$, and the convolution of $\hat{S}_0$ with $\hat{h}$ is subtracted from the original DC-signal $S_{DC}$, leading to $\hat{S}_m$.

The estimated convolution kernel $\hat{h}$ was then extracted by deconvolving the original DC-signal $S_{DC}$ with $\hat{S}_0$. With $\star^{-1}$ as deconvolution operator, equation [3] leads to:

$$S_{DC} \star^{-1} \hat{S}_0 = \hat{S}_m \star^{-1} \hat{S}_0 + \hat{h} \star \hat{S}_0 \star^{-1} \hat{S}_0 \qquad [4]$$

To solve this problem, a deconvolution algorithm was adapted by fitting the general model $\hat{S}_0 \star \tilde{h}$ onto the signal $S_{DC}$ in the least squares sense. $\hat{S}_0$ being known and the coefficients of $\tilde{h}$ being the fitting parameters, the optimization process leads to the equality between the kernel $\tilde{h}$ and $S_{DC} \star^{-1} \hat{S}_0$. As $\hat{S}_m$ and $\hat{S}_0$ are not physically correlated, the solution for $\hat{S}_m \star^{-1} \hat{S}_0$ out of the deconvolution algorithm is expected to be very close to a null vector. Moreover, $\hat{S}_0 \star^{-1} \hat{S}_0$ yields a solution very close to a Dirac. Hence, equation [4] is simplified to $\tilde{h} \approx \hat{h}$, which provides the estimation of $h$ we want to extract. $\hat{S}_0$ and $\hat{h}$ being known, $S_{DC}$ can be corrected with SNIF by extracting the motion component through $\hat{S}_m = S_{DC} - \hat{S}_0 \star \hat{h}$ (Figure 1e).

## Methods

Thoracic MR acquisitions were performed on a 3.0 T scanner (Signa PET/MR, GE Healthcare, Waukesha, WI, USA) on a freely breathing healthy human volunteer lying supine in a 14-channel spine array and a 16-channel thoracic array for signal reception, equipped with a pneumatic abdominal respiratory belt and a photoplethysmograph (PPG) on the finger to output surrogate signals for respiratory and cardiac motions respectively. Tissue RF excitation was performed with the body coil of the MR scanner. The study was undertaken according to general clinical research guidelines and regulations. It was approved by our local ethics committee (CEA-SHFJ Scientific Board). Informed consent was obtained from the volunteer.

A 3D-radial ('kooshball') UTE sequence with a random order of spoke orientations was implemented with flip angle $\alpha = 3°$, $TE = 100$ μs, $TR = 2$ ms, FOV $= (32 \times 32 \times 22.4)$ cm$^3$, Voxel size $= (1.3 \times 1.3 \times 1.3)$ mm$^3$, full sampling, 3000 dummy excitations, RF spoiling and ramp-sampling. It resulted in a first acquisition of $N_{\text{spokes}} = 150{,}806$ for a total scan time of 5 min 2 s. Spoiling gradients were inserted in between spokes with an increased $TR = 3.8$ ms to yield a second acquisition of $N_{\text{spokes}} = 161{,}062$ for a total scan time of 10 min.

A complete set of spokes was initially defined to cover k-space uniformly. The k-space trajectory was then computed by $N$spokes permutations of every spoke with another randomly-selected spoke to build a complete random order of the spokes.

In the experiment, $S_{DC}$ was the complex value of the first acquired data point for each spoke at a given reception channel. The analysis was carried out separately on real and imaginary parts for each of the 30 coil channels. For the first acquisition $TR = 2$ ms, $N_{\text{rep}} = 50$ previous repetitions were considered for the complex convolution kernel $\hat{h}$, leading to $\hat{h} = [\hat{h}_{50} \ldots \hat{h}_3\, \hat{h}_2\, \hat{h}_1\, \hat{h}_0]$, which was sufficient in the current experimental conditions as checked with $N_{\text{rep}} = \{25,50,75,100\}$. Thereafter, this choice was used to correct for the spin-history artifact for the second acquisition at $TR = 3.8$ ms.

The efficiency of SNIF was evaluated on the real and imaginary DC signal temporal spectra for $S_{DC}$ and $S_m$, respectively before and after processing. For most of the reception channels, the DC complex spectra exhibited motion-related peaks of the first three respiratory harmonics and the first two cardiac harmonics which were confirmed by the temporal spectra of the simultaneously acquired respiratory and cardiac tracks. The root mean square (rms) envelope of the spectra was computed and the associated baseline was removed before running a standard peak analysis under Matlab®. The frequencies, the prominences, and the widths of the peaks associated with the physiologic motions were extracted for every reception channel. The DC-signal frequencies were compared to respiratory and cardiac harmonic frequencies of the respiratory belt and PPG signals. The $S_{DC}$ and $S_m$ peak prominence and width ratios were calculated for every motion harmonic in each reception channel.

## Results

Figure 2 shows a twenty-second close-up of a typical DC-signal extracted from one of the 30 reception channels before and after the correction of off-center and spin-history artifacts by SNIF. This reception channel was chosen as it was able to measure signal variations related to both respiratory and cardiac motions. The DC-signal is clearly improved after the application of SNIF. The noticeable reduction of the amount of apparent noise enables a better depiction of the respiratory (long red arrow) and cardiac (short red arrow) motion-related oscillations.

Figure 3 shows the imaginary DC-signal associated spectra obtained before and after SNIF with the physiologic spectrum from the respiratory belt as a reference. The latter was scaled down by a factor of 35. Three of the thirty available reception channels are presented: a. Coil 2, centered onto the posterior neck; b. Coil 20, centered onto the anterior thorax, in front of the heart; c. Coil 28, centered onto the diaphragm side, just below the thoracic cage (see Figure 4 for coil element positioning). In coil 2, the DC-signal is rather immune to physiologic motions that occur below, and spectra in a. show peaks essentially hidden by the noise in spite of the rms smoothing. In coil 20, five peaks clearly point out the first three respiratory harmonics and the first two cardiac harmonics. Their positions match the physiologic track spectra. Coil 20 is sensitive to both respiratory and cardiac motion whereas coil 28 is mostly sensitive to respiratory motion. The spread of dots, which represent raw spectral data in Figure 3, already highlights the SNR gain by SNIF processing. For every reception channel, we observe both the global energy conservation for each motion harmonic and the decrease of the background noise around each peak.

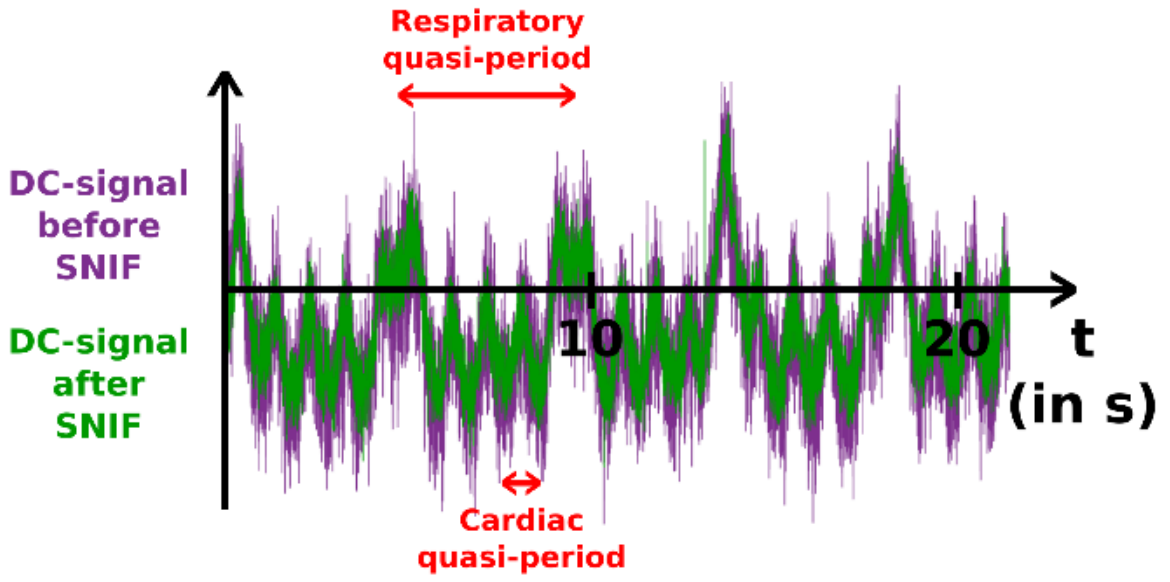

**Figure 2:** Close up over twenty seconds for one reception channel of the DC-signal real part before (magenta) and after SNIF (green) with both off k-space center and spin-history artifacts corrections with $N_{\text{rep}} = 50$.

Figure 4 shows the ratios of the peak prominences, $p = P_m/P_{DC}$, and widths, $w = W_m/W_{DC}$, before and after SNIF for the respiratory, $p_{\text{r1st}}$ and $w_{\text{r1st}}$, and cardiac, $p_{\text{c1st}}$ and $w_{\text{c1st}}$, first harmonics extracted from the first dataset at $TR = 2$ ms. The ratios are mapped onto the 14-channel spine coil array and the 16-channel thorax coil array. The peak prominences are enhanced ($\langle p_{\text{r1st}} \rangle > 1$ and $\langle p_{\text{c1st}} \rangle > 1$) while the peak widths are reduced ($\langle w_{\text{r1st}} \rangle < 1$ and $\langle w_{\text{c1st}} \rangle < 1$), namely and, which favors peak selectivity.

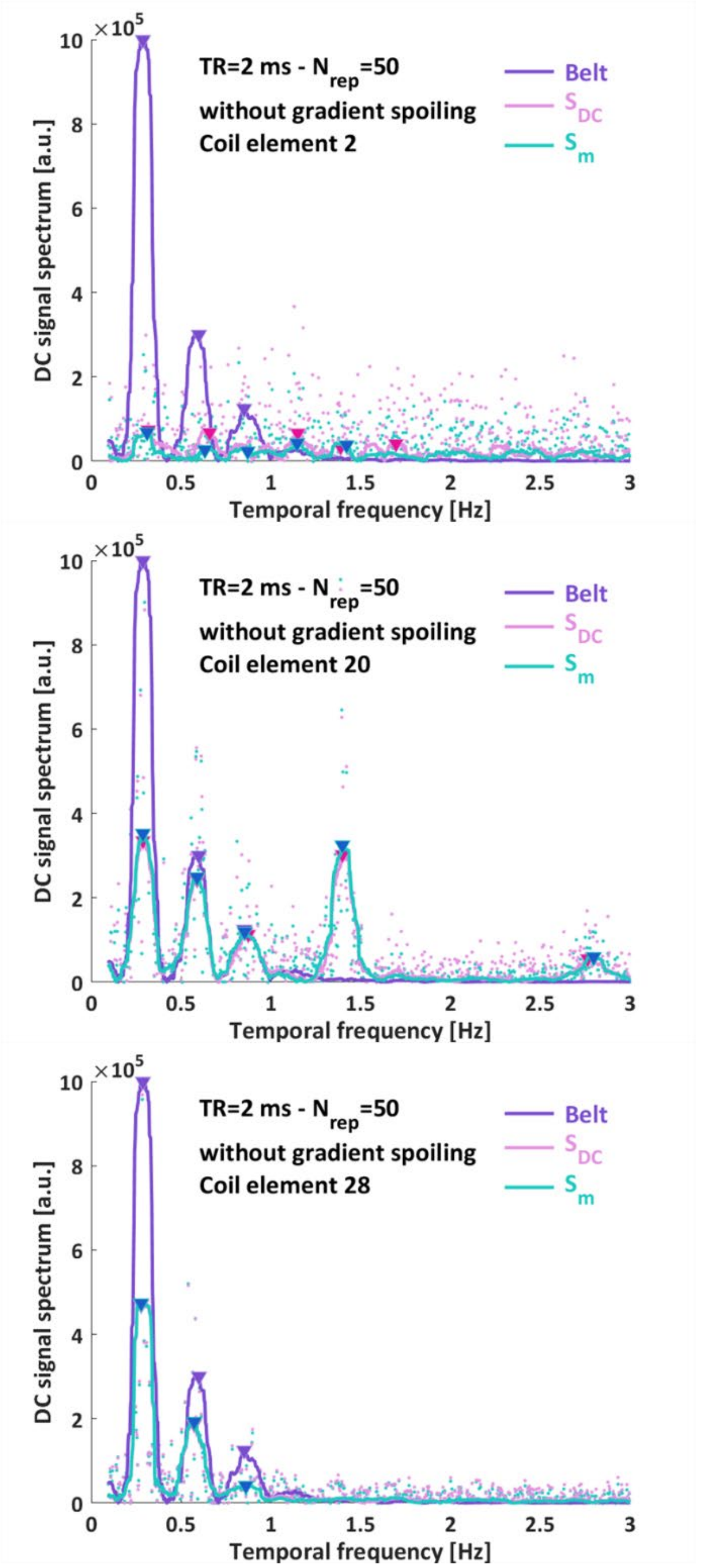

**Figure 3:** Imaginary DC-signal associated spectra obtained before (pink) and after (cyan) SNIF for a. Coil 2, b. Coil 20, and c. Coil 28 (see Figure 4 for positioning). Dots are detrended raw spectral data and lines are rms spectral envelopes. The respiratory track spectrum from the belt (purple) is scaled down by a factor of 35. Complete sets of imaginary, real, and magnitude DC-signal spectra are provided for every reception channel as supplementary material.

Cardiac motion is barely detected anywhere but close to the coil elements 20-21 which faced the subject's heart. Elsewhere, cardiac peaks are difficult to extract as indicated by the white elements, for which peaks could not be detected. Thus the posterior cardiac values, extracted from the spine coil array, must not be taken into account as they have little to do with the cardiac motion. On the contrary, the anterior cardiac values are advantageously enhanced by SNIF with an average 22% improvement in the peak prominence, which is strengthened by a 4% peak thinning throughout the thorax coil elements. When the first harmonics are extracted from the second dataset at $TR = 3.8$ ms, the mean ratios are very close to one and SNIF correction is not noticeable.

Figure 5 shows peak frequencies in Hertz of the first three respiratory harmonics mapped onto the 16-channel thorax coil array before and after SNIF for the first dataset at $TR = 2$ ms. Mean and standard deviation frequency values are indicated under each map. After SNIF, the values ($(0.29 \pm 0.01)$ Hz, $(0.58 \pm 0.01)$ Hz, $(0.88 \pm 0.07)$ Hz) better match the reference values extracted from the respiratory track spectrum: $f_{\text{r1st}} = 0.285$ Hz, $f_{\text{r2nd}} = 0.574$ Hz, and $f_{\text{r3rd}} = 0.869$ Hz. The standard deviation among the processed coil DC-signal is also reduced after SNIF. For the second dataset at $TR = 3.8$ ms, the mean values ($(0.33 \pm 0.01)$ Hz, $(0.66 \pm 0.02)$ Hz, $(1.04 \pm 0.12)$ Hz) remain unchanged after SNIF. They already matched fairly well the reference values extracted from the respiratory track spectrum: $f_{\text{r1st}} = 0.332$ Hz, $f_{\text{r2nd}} = 0.679$ Hz, and $f_{\text{r3rd}} = 1.059$ Hz.

## Discussion and Conclusions

The underlying theory of SNIF is built on several strong assumptions. The first one is the existence of an off-center component. When represented in a $(\theta, \varphi)$ coordinate system, the acquired data show a coherent but non-uniform pattern that stands out of what can be named as noise (Figure 1b). On the one hand, this noise can be easily understood as physiologic motions and spin-history effects lose any apparent coherence when represented in $(\theta, \varphi)$ because of the quasi-random order of spoke orientations. On the other hand, the subject's MR image is experimentally spatially continuous, so is its spatial Fourier representation. Hence, a smooth pattern can be observed and extracted out of the k-space by low-pass filtering the non-coherent components in $(\theta, \varphi)$.

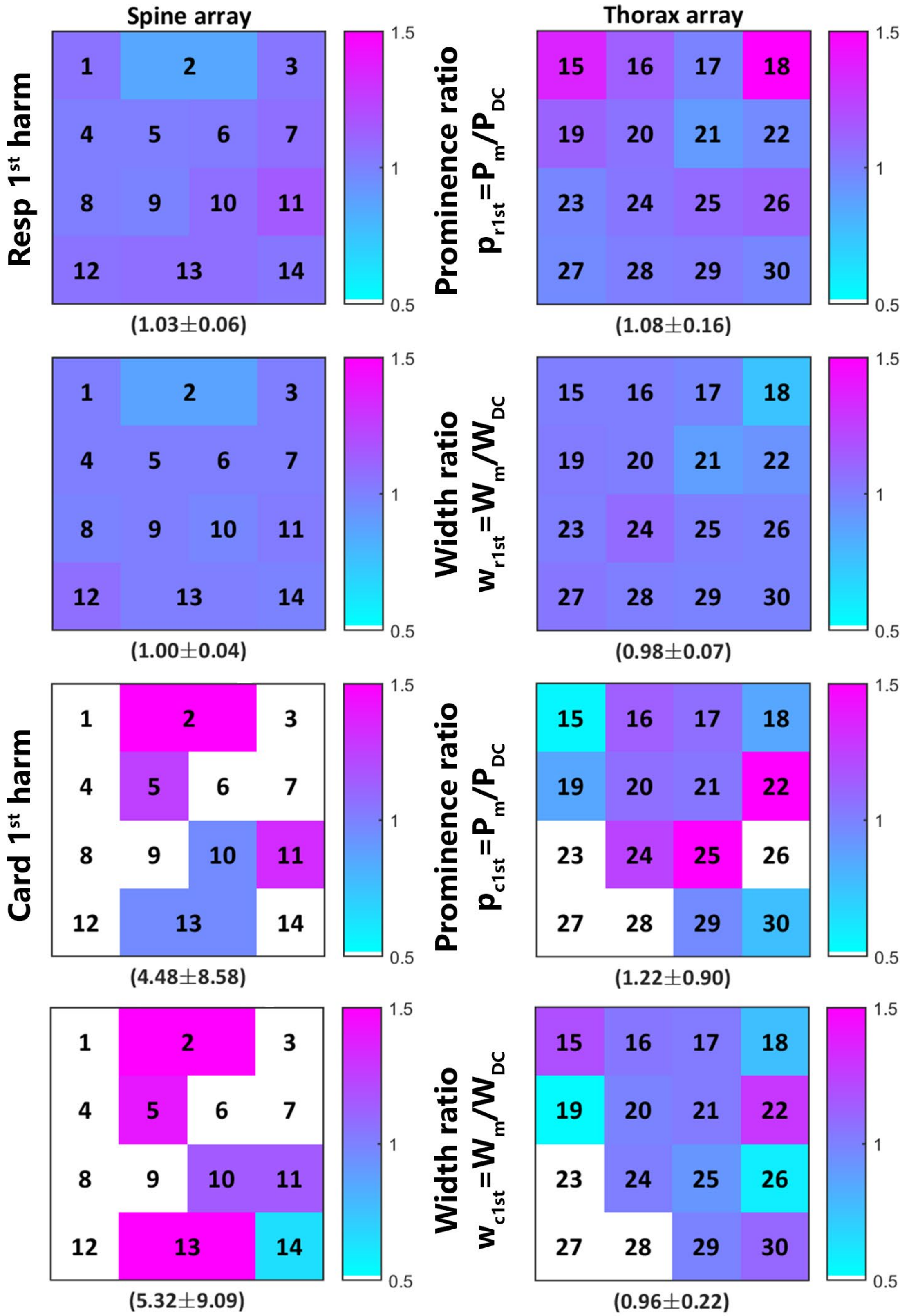

**Figure 4**: Ratios of peak prominences and widths before and after SNIF for the respiratory (top rows) and cardiac (bottom rows) first harmonics mapped onto the 14-channel spine coil array (left) and the 16-channel thorax coil array (right). Mean and standard deviation ratios are indicated under each map. The 30 coil elements are numbered and presented with respect to a coronal view. Peaks in white elements could not be determined and analyzed.

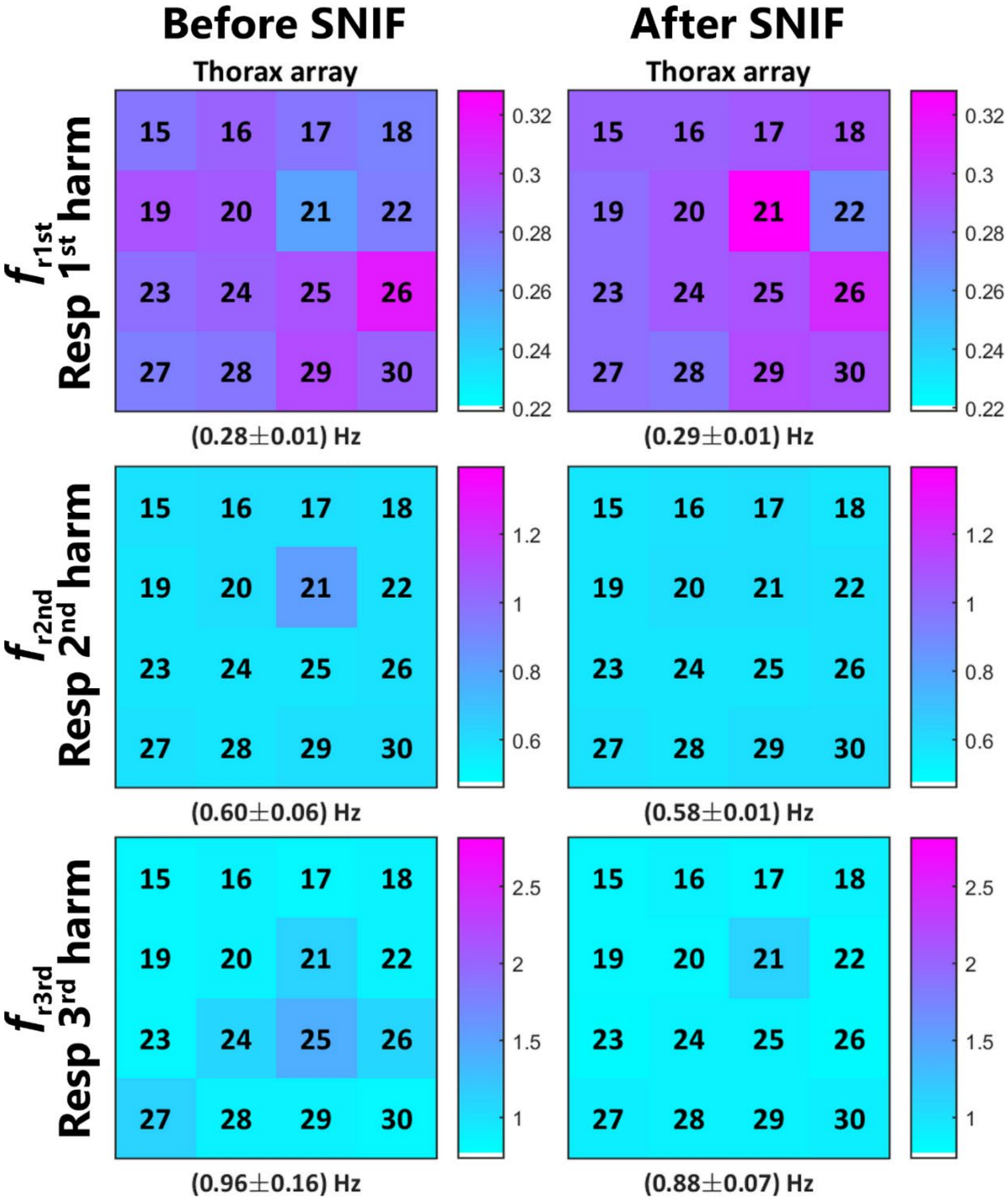

**Figure 5**: Peak frequencies in Hertz of the first three respiratory harmonics mapped onto the 16-channel thorax coil array, (left) before and (right) after SNIF. Mean and standard deviation frequency values are indicated under each map. For reference, upon the respiratory track spectrum, the frequencies of the first three harmonics are $f_{\mathrm{r1st}} = 0.285$ Hz, $f_{\mathrm{r2nd}} = 0.574$ Hz, and $f_{\mathrm{r3rd}} = 0.869$ Hz. The 16 thorax coil elements are numbered and presented with respect to a coronal view. Scales match the range of each harmonic frequency.

The second assumption, which states the influence of spin-history, is ascertained by the outcomes of the deconvolution process yielding the estimation $\hat{h}$ of $h$. The coefficients of $\hat{h}$ at negative indexes ($\hat{h}_{-1}$, $\hat{h}_{-2}$, ...), which could support a backward dependence of $S_{DC}$ on future values of $S_0$, are experimentally negligible (Figure 6) with respect to the coefficients of $\hat{h}$ at positive indexes ($\hat{h}_1$, $\hat{h}_2$, ...). Therefore, the current magnetization state is effectively influenced by past magnetization states but not by future ones, in accordance with the causality principle.

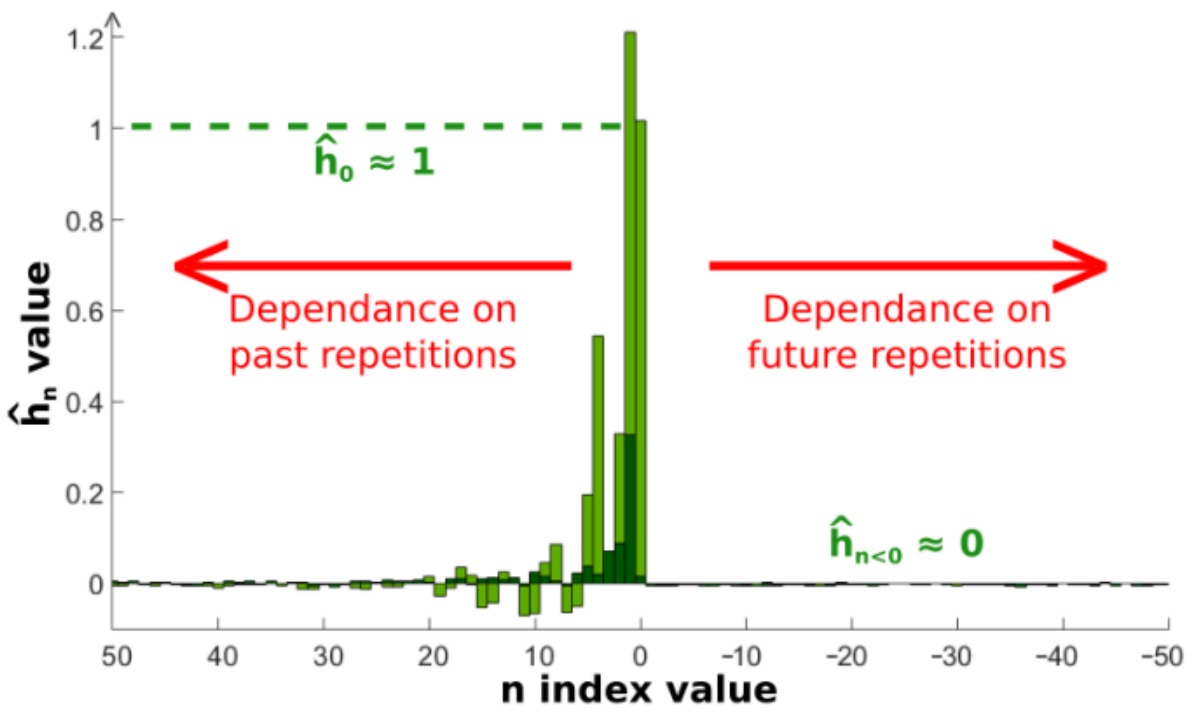

**Figure 6:** Representation for one reception channel of the values estimated for $\hat{h}$ by deconvolving the original DC-signal $S_{DC}$ with $\hat{S}_0$, with the real part of $\hat{h}$ represented in light green and the imaginary part in dark green. For the sake of clarity, in order to verify the coherence of what is computed, the correlation coefficients $\hat{h}_n$ were estimated for positive $n$ values (left part of the graph), corresponding to a correlation with the past repetitions (spin-history effect), but also for negative n values (right part of the graph), which would correspond to a non-physical correlation with future repetitions. First, we observe that $\hat{h}_0$ is very close to 1, which confirms the right extraction of $\hat{S}_0$. Second, we observe non-zero complex $\hat{h}_n$ values for positive $n$, confirming the existence of a spin-history effect. These weights globally fade as $n$ increases. Furthermore, $\hat{h}_n$values are close to zero for negative $n$, which is physically expected: at a given repetition, and considering the randomized order of spoke orientations, the transverse magnetization state cannot be correlated with transverse magnetization states reached later during future repetitions.

The third assumption considers that the spin-history artifact follows a linear model based on the convolution kernel $h$. This is reasonable since the spin-history effect can be seen as a cumulative perturbation, which depends on the transverse magnetization states of preceding sequence repetitions. Each transverse magnetization state is assumed to be proportional to the $S_0$ values at the same repetition times, with $h$ exhibiting the coefficients of proportionality between the $S_0$ values and the correction terms for the $N_{\text{rep}}$ preceding repetitions taken into account to estimate and correct the spin-history artifact. When applied to data acquired with effective gradient spoiling, with $TR = 3.8$ ms, SNIF has no measurable effect as the spin history is erased every $TR$.

Finally, a fourth assumption is made by introducing the motion component $S_m$ as an additional term in the measured DC-signal. It is straightforward to justify when both off-center and spin-history artifacts are neglected. In this case, the mean intensity $S_{DC}(t)$ of an image that would be acquired instantaneously at $t$ is the sum of i) the mean image intensity without motion $\langle S_0(t)\rangle = S_0$ and ii) the motion correction term $S_m(t)$: $S_{DC}(t) = S_0 + S_m(t)$. In this situation, the last term characterizes the variation of $S_{DC}$ around $S_0$ related to the subject's motion. Experimentally, when off-center and spin-history artifacts cannot be neglected, the mean image intensity without motion $S_0(t)$ is not constant anymore but evolves with the off-center jitter and cumulates the weighted preceding transverse magnetizations through convolution with $h$, leading to equation [2]. [2] can be simplified to [3] and the spin-history effect is correctly estimated as long as $T \gg N_{\text{rep}}TR$, namely as long as the motion characteristic time $T$ is much longer than the spin-history characteristic time, which is fulfilled here with $N_{\text{rep}} = 50$, $TR = 2$ ms or $TR = 3.8$ ms, $T_{\text{cardiac}} \geq 500$ ms, and $T_{\text{resp}} \geq 3000$ ms for a healthy adult. Moreover, in Figure 6, we retrospectively observe that the value $N_{\text{rep}} = 50$ is experimentally enough as the $\hat{h}_{20\leq n\leq 50}$ coefficients are very low compared to $\hat{h}_{0\leq n\leq 10}$ values in terms of both real and imaginary parts. For that reason, looking back in the past for $N_{\text{rep}} > 50$ neither improves the correction nor yields a better self-navigator, but it might add unwanted biases and would unnecessarily increase the post-processing time and complexity. This time could even be halved here by halving $N_{\text{rep}}$ as no deterioration of the correction was observed down to $N_{\text{rep}} = 25$.

The correction of self-navigator biases was studied in several papers [16,17]. Those methods are more general, in the sense that they remove the artifacts from both the self-navigator and the final images. Also, they are usually not limited to quasi-random trajectories. However, to our knowledge, SNIF is the only one to enable the correction of both the off-center and the spin-history artifacts. Moreover, the underlying model proposed by these other methods is often more complicated (the relationship between reception channels needs to be considered, for example), and the principle and usability of the correction pipeline are usually not as simple as for SNIF (a calibration scan may be required, for example). To reduce spin-history artifacts, phyllotaxis trajectories may be considered [18]. However, these more advanced trajectories are not as adapted as those featuring a quasi-random sampling for retrospective motion correction. This is particularly true with retrospective gating methods because of the non-uniform sampling of k-space resulting from each retrospective gate in non-quasi-random trajectories.

SNIF is thus robust and should be applicable to 2D or 3D radial MR acquisition, whatever acquisition parameters are selected, provided quasi-random order of spoke orientations is designed and the condition $T \gg N_{\mathrm{rep}}TR$ is verified, which is the case for many standard pulse sequences, and especially for the UTE sequence performed here. In an ideal situation where there would be neither off-center nor spin-history artifacts, where $S_0$ would then be time-independent and $h$ would be a Dirac, SNIF would not alter the DC-signal. In other words, provided the DC-signal is blurred by off-center and spin-history effects and in the absence of bulk motion, irregular breathing or coil-sensitivity changes, SNIF is expected to improve the extraction of self-navigators from the DC-signal. SNIF is also expected to provide a signal more closely related to the lung and heart motion fields inside the thorax because the sources of biases are physically understood, modelled and filtered with SNIF. If needed, the residual noise experimentally observed (Figure 2) could additionally be removed with a more standard method (linear bandpass filtering, for example). Ultimately, SNIF outcomes might advantageously be used for motion correction in MRI or hybrid PET-MRI image reconstruction.

## Acknowledgements

PET/MR platform affiliated to the France Life Imaging network (grant ANR-11-INBS-0006). This work was supported in part by NIH grant R01HL136965.

## Competing interests

Dr. Tanguy Boucneau and Dr. Brice Fernandez are currently employees of GE Healthcare. The other authors declare no conflicts of interest.

## Additional information

Supplementary information is available for this paper.